\documentstyle[times,psfig,graphics,astrobib,amssymb]{mn2e}

\newcommand{\be}{\begin{equation}}
\newcommand{\e}{\end{equation}}
\newcommand{\bear}{\begin{eqnarray}}
\newcommand{\ear}{\end{eqnarray}}
\newcommand{\nline}{\nonumber \\}
\newcommand{\f}{\frac}
\newcommand{\de}{{\rm d}}
\newcommand{\del}{\partial}

\begin{document}

\title[A theoretician's analysis of the supernova data]
{A theoretician's analysis of the supernova data
and the limitations in determining the nature 
of dark energy}
\author[Padmanabhan \& Choudhury]
{T. Padmanabhan\thanks{E-mail: nabhan@iucaa.ernet.in},
T. Roy Choudhury\thanks{E-mail: tirth@iucaa.ernet.in}\\
IUCAA, Ganeshkhind, Pune, India 411 007}

\maketitle

\date{\today}

\begin{abstract}
Current cosmological observations show a strong signature of 
the existence of a 
dark energy component with negative pressure.  The most obvious
candidate for this dark energy is the cosmological constant (with 
the equation of state $w_X = p/\rho = -1$), which, however, 
raises several theoretical difficulties. This has led to 
models for  dark energy component 
which evolves with time. We 
discuss certain questions related to the 
determination of the nature of 
dark energy component from observations of high redshift supernova.
The main results of our analysis are: (i) Even if the precise 
value of $w_X$ is known from 
observations, it is {\em not} possible to determine the nature of 
the unknown dark energy source using only kinematical
and geometrical measurements. We have given explicit 
examples to show that different types of sources 
can give rise to a 
given $w_X$. (ii) Although 
the full data set of supernova observations (which are currently available) 
strongly rule out models without dark energy, 
the high ($z > 0.25$) and low ($z < 0.25$) redshift 
data sets, individually, admit decelerating models with 
zero dark energy. Any possible evolution 
in the absolute magnitude of the supernovae, if detected, 
might allow the decelerating models to be consistent with 
the data. (iii) We have introduced two parameters, which 
can be obtained entirely from theory,  
to study the sensitivity of the luminosity distance on $w_X$.
Using these two parameters, we have argued that although 
one can  determine the present value of $w_X$ accurately from 
the data, one cannot constrain the evolution of $w_X$. 
\end{abstract}

\section{Introduction}

  One of the equations governing the dynamics of a Friedman universe, 
  $(\ddot a/a) = - (4\pi G/3)(\rho+3p)$, implies that the universe will
  accelerate ($\ddot a >0$) if $(\rho + 3p) <0$. The analysis of high-redshift
  supernova data \cite{rfc++98,pag++99,riess00} seems to suggest that
  $\ddot a$ is indeed positive thereby requiring $(\rho + 3p) <0$.
  This condition requires at least $\rho$ or $p$ to be negative. Cosmologists
  have not yet become desperate enough to suggest  $\rho<0$; 
  but there was remarkably low resistance in the community to accepting
   the existence  of a constituent with  $w \equiv (p/\rho)<-(1/3)$.
   This was, to a great extent, facilitated by the fact that cosmologists
   have long since toyed with the existence of a non-zero cosmological constant
   with the stress-tensor $T^i_k = \rho_{\Lambda} \delta^i_k$ 
corresponding to $w_{\Lambda}=-1$.
   In fact,  much before the arrival of supernova data on the scene, 
there were 
   indications that the universe actually may be populated by a energy density
   which is very smoothly distributed over large scales. A very clear 
   argument to this effect was given based on the analysis of APM data in 1990
   \cite{esm90}. Later, the first analysis of the COBE data in 1992 
indicated that the 
   shape of the standard cold dark matter (SCDM) spectrum needs to 
be modified and the existence
   of a smoother energy distribution, like the cosmological constant will be 
   required \cite{pn92,ebw92}. 
An analysis of a host of observations available in 1996 was
   used to suggest that the data supports 
the existence of a non-zero cosmological
   constant \cite{bpn96}.
   Given this background, it was not surprising 
that the cosmologists were not too shocked
   by the supernova data which became available from 1998 onwards. 
Currently, there are also other observations, like those of the age of the 
universe, gravitational lensing surveys etc., which 
strongly suggest the presence of a positive cosmological constant 
[see \citeN{padmanabhan02c} for a comprehensive review and 
references].

   There are, however, well known deep theoretical 
problems with the existence of a
   non-zero cosmological constant $\Lambda$ with a magnitude 
of about $\Lambda (G\hbar/c^3)
   \approx 10^{-123}$. [These are well documented in the literature 
and will not be discussed
   here; for a recent review see \citeN{padmanabhan02c}.]
This has prompted  a host of activity in which 
   one looks for a dark energy component in the universe 
with $w<-(1/3)$ which is different
   from cosmological constant [see \citeN{padmanabhan02c} and references cited therein].
   If $w_X(a)$ denotes the time dependent equation of state 
parameter for an unknown
   dark energy component in the universe, then the 
equation of motion $\de (\rho_X a^3) =  w_X \rho_X
   \de a^3$ integrates to give
   \begin{equation}
    \rho_X(a) = \rho_X(a_0) 
\exp \left\{- 3 \int_{a_0}^a \frac{\de a'}{a'} 
[1 + w_X(a')] 
  \right\}
  \label{evolenergy}
   \end{equation} 
   If $w_X$ is not identically equal to $-1$, then the dark 
energy density will evolve with time (even
   if $w_X$ is a constant). This suggests the 
possibility that the cosmological ``constant'' can be time 
   dependent thereby alleviating some of the difficulties which 
arises in the case of $w_X=-1$.
   Several such models (usually based on scalar fields) have been 
constructed in the literature,
   all of which, generically have a time dependent $w_X(a)$. The 
introduction of such an ad-hoc,
   time dependent \emph{function} into cosmology, of course, 
takes away much of the 
   credibility; nevertheless, this 
issue needs to be settled ultimately by observations and not
   by aesthetics. (If the observations forces us into a solution 
which is unaesthetic by current standards,
   we will either conveniently change the standards or live with it!) 
   
   In this paper, we discuss certain questions related to the 
determination of the nature of 
   dark energy component from observations related to supernova. This 
issue has been 
   studied in great detail by several groups and our contribution 
in this paper will border on
   pedagogy and will present a particular point of view.  
In the next section we briefly recall
   and stress some inherent theoretical degeneracies in these analysis.  
In Section \ref{sndata}
   we reanalyze the currently available supernova data in order to focus 
attention on some 
   key elements. (These results exist in alternate forms in published 
literature but we believe
   our analysis brings out some features clearly.) Based on the 
lessons learnt in this section,
   we carry out a similar analysis for possible future supernova 
observations and point out
   what can and cannot be achieved. We have intentionally kept  
the data analysis at a fairly
   simple level (indicated in the title by `a theoretician's analysis'). 
We subscribe to the point of 
   view that any result which cannot be revealed by a simple 
analysis of data, but arises
   through a more complex statistical procedure, is inherently 
suspect and a conclusion as 
   important as the existence of dark energy with 
$p<0$ should pass such a test.

\section{Theoretical degeneracies in the Friedmann model}
\label{theodegen}

We begin by recalling and stressing some inherent limitations 
which exists in all 
attempts to probe the universe through geometrical measures. 
The assumption of isotropy and homogeneity implies that 
the large scale geometry of the universe can be described 
by a metric of the form
\be
\de s^2 = \de t^2 - a^2(t) \de {\bf x}^2
\e
where 
$\de {\bf x}^2$ 
denotes the line element of the 
three-dimensional space in comoving coordinates. 
In any range of time during which $a(t)$ is a monotonic 
function of $t$, one can use $a$ (or equivalently,  
redshift $z = (a/a_0)^{-1}-1$) as a time coordinate. The metric 
is then
\be
\de s^2 = \f{1}{H^2(a)} \left(\f{\de a}{a}\right)^2 - a^2 \de {\bf x}^2
=\f{1}{(1+z)^2} \left[\f{\de z^2}{H^2(z)} - \de {\bf x}^2\right]
\e
where $H(z) = \dot{a}/a$ is the Hubble parameter.

This equation allows us to draw the following  conclusion: The only 
non-trivial metric function in a Friedmann universe is the 
function $H(z)$ (besides the curvature of the spatial 
part of the metric). Hence, any kind of observation 
based on geometry of spacetime, however complex it may be, 
will not allow us to determine anything other than 
this single function $H(z)$. 
Since Friedmann equations relate $H^2(z)$ to the \emph{total} 
energy density in the universe (assuming 
that the curvature of the spatial 
part of the metric
is known from independent 
observations or fixed by some theoretical prejudice), the best we can 
do from any geometrical observation is to determine the total energy density
of the universe at any given $z$. It is not  possible to determine the energy
densities of individual components from any geometrical observation. 

More explicitly, when  several 
non-interacting sources are present in the universe, the Friedmann
equations give
\bear
H^2(z) &=& H_0^2 \left[\sum_{\alpha} \Omega_{\alpha} \exp\left 
\{3 \int_0^z \f{\de z'}{1+z'} [1 + w_{\alpha}(z')]\right\} 
\right.\nline
&+& \left. \Omega_k (1+z)^2 \right], \nline
\Omega_k &=& 1 - \sum_{\alpha} \Omega_{\alpha},
\ear
where $\Omega_{\alpha}$ is the density parameter for  the $\alpha$-th 
component (like radiation, matter, cosmological constant etc.) 
and $w_{\alpha}(z)$ is the corresponding equation of state parameter. For 
example, 
non-relativistic matter ($\Omega_m$) has $w_m = 0$, while 
a non-evolving 
cosmological constant ($\Omega_{\Lambda}$) has $w_{\Lambda} = -1$.
From the above equation, it is clear that one needs to know the 
$\Omega_{\alpha}$ and  $w_{\alpha}(z)$ of the 
$N-1$ components [along with
geometrical measurements giving  $H(z)$] 
in order to determine the contribution
of the $N$-th component. 
If we take $\Omega_k =0$ corresponding to the flat universe,
and assume that there is only dust-like 
matter and dark energy present in the universe,
then we need to know $\Omega_m$ of 
non-relativistic dust-like matter and 
$H(z)$ to get a handle on $\Omega_X$ of the dark energy. 
Unfortunately, even after making these assumptions,
we are plagued by the uncertainties in 
$\Omega_m$ which is currently estimated to
be anywhere between 0.2 and 0.35. 
As has been noted several times in the literature
(and as we shall emphasize in Section \ref{evoldarken}), 
this is a fairly strong degeneracy.
 
There is another --- and from a theoretical point of view more serious ---
degeneracy which seems to have been inadequately stressed in the literature.
Let us assume that the universe is made of two components: 
$\rho_{\rm kn}(a)$, which is known from
independent observations and a component $\rho_X(a)$ which is not known. 
From the Friedmann equation, it follows that
\begin{equation}
\frac{8\pi G}{3}\rho_X(a)=H^2(a)[1-\Omega_{\rm kn}(a)];
\quad \Omega_{\rm kn}(a)\equiv\frac{8\pi G\rho_{\rm kn}(a)}{3 H^2(a)}
\label{findingrhox}
\end{equation}
Taking a  derivative of $\ln \rho_X(a)$  and using (\ref{evolenergy}), 
it is easy to obtain the relation
\begin{equation}
w_X(a)=-\frac{1}{3}\frac{\de}{\de \ln a}
\ln\left\{[1-\Omega_{\rm kn}(a)] H^2(a) a^3\right\}
\label{findingw}
\end{equation}
If geometrical observations of the universe give us $H(a)$ 
and other observations give us $\rho_{\rm kn}(a)$ then
one can determine $\Omega_{\rm kn}(a)$ and thus $w_X(a)$. 
This is the best uniform background cosmology can do for
us. As far as cosmological model building is concerned, 
one is usually interested in the features of the 
dark energy ``fluid'', and it 
is usually sufficient to 
know $w_X(a)$ and the speed of sound of the fluid. However, 
theoretical physicists would like to know something more 
about the nature of the dark energy ``fluid'' -- in particular, 
they would find it useful to determine the Lagrangian for the fluid. 
(For example, in the case of inflation, 
a considerable amount of effort is being spent to understand the 
form of the potential for the scalar field.) In this paper, we 
shall take the viewpoint of a theoretical physicist, where the nature of the 
dark energy is defined by its Lagrangian.

\begin{figure*}
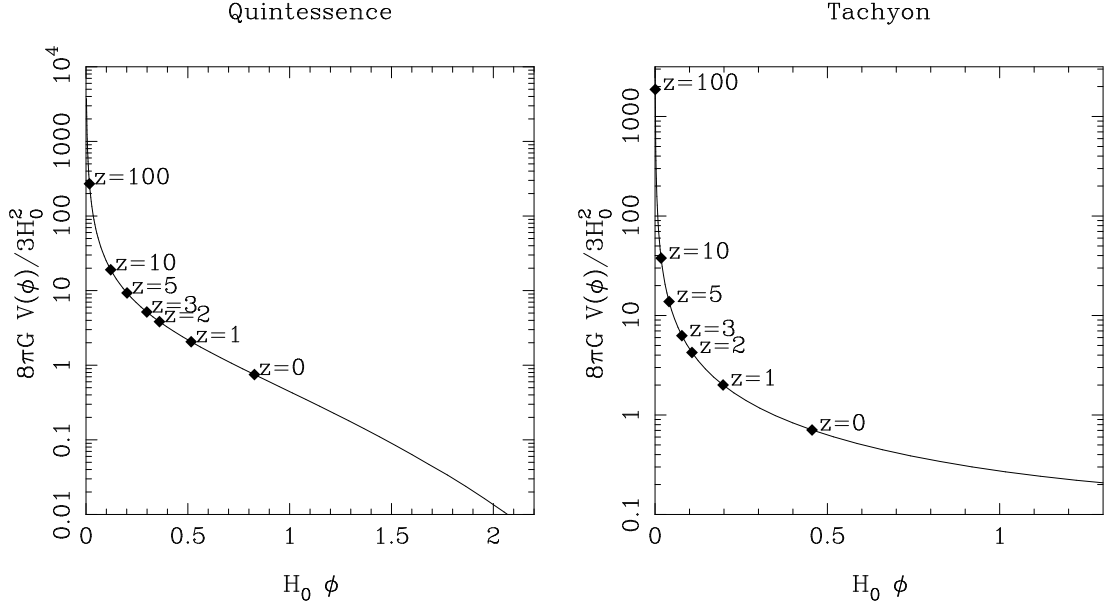

\begin{center}
\rotatebox{270}{\resizebox{0.45\textwidth}{!}{\includegraphics{qpot.ps}}}
~~~~~~~\rotatebox{270}{\resizebox{0.45\textwidth}{!}{\includegraphics{tpot.ps}}}
\caption{
The scalar field potentials for the quintessence (left frame) and tachyonic 
(right frame) fields which are required to produce 
the dark energy equation of state $w_X(a) = w_0 - w_1 (a -1)$. The 
potentials are plotted for the parameter values 
$\Omega_m = 0.3, \Omega_X = 1 - \Omega_m = 0.7, 
w_0 = 0.5, w_1 = -0.1$ using equations 
(\ref{voft}) -- (\ref{finalone}). 
Since the equations give the potential $V(\phi)$ in the parametric form 
[$V(z)$, $\phi(z)$], each point on the $V - \phi$ curve can be labelled
by the corresponding value of redshift, $z$. 
We have pointed some particular redshifts in both the curves.
It is clear that 
the scalar field 
starts from a large value of the potential, rolls down as time 
progresses and the density contributed by the potential 
reaches a value $\sim \Omega_X$ at $z = 0$.}
\label{vphi}
\end{center}
\end{figure*}
Now, if a geometrical 
observation in future suggest that $w_X(a)$ is 
not identically equal to $-1$, then
the question arises as to what is the nature of this dark energy field. 
It seems virtually impossible
that its nature can be determined from laboratory experiments 
(unlike, for example, WIMPS which
might constitute dark matter). However, knowing $w_X(a)$ 
is grossly inadequate for determining the 
physical nature of dark energy. For example, even 
if one makes another gigantic leap of faith
and assumes that the dark energy arises from a scalar field, 
it is possible to come up with
scalar field Lagrangians of different forms leading to same $w_X(a)$ 
[see \citeN{padmanabhan02}].  To illustrate this point,  we shall 
discuss two possibilities :
\be
L_{\rm quin} = \f{1}{2} \del_i \phi \del^i \phi - V(\phi), ~~
L_{\rm tach} = - V(\phi) \sqrt{1 - \del_i \phi \del^i \phi}
\e
Both these Lagrangians involve one arbitrary function $V(\phi)$. 
The first one, $L_{\rm quin}$, which is a natural 
generalization of the Lagrangian for a non-relativistic 
particle [$L = \dot{q}^2/2 - V(q)$], is usually called 
quintessence. Similarly, the second one, $L_{\rm tach}$, is 
a generalization of the Lagrangian for a relativistic 
particle [$L = -m \sqrt{1 - \dot{q}^2}$], and is usually called 
a tachyonic Lagrangian. [It arises naturally in the string 
theoretical context; see \citeN{sen02c}; \citeN{sen02}; \citeN{sen02b}.
For a discussion of the cosmological 
 aspect of this Lagrangian, see \citeN{pc02} and references therein.] 
When these
Lagrangians act as sources in Friedmann universe, they 
are characterized by density and equation of state 
parameters given by
\be
\rho_{X, {\rm quin}} = \f{1}{2} \dot{\phi}^2 + V, ~~
w_{X, {\rm quin}} = \f{1 - (2 V/\dot{\phi}^2)}{1 + (2 V/\dot{\phi}^2)}
\e
and
\be
\rho_{X, {\rm tach}} = V \sqrt{1 - \dot{\phi}^2}, ~~
w_{X, {\rm tach}} = \dot{\phi}^2 - 1,
\e
respectively.

Since both the Lagrangians have one undetermined 
function $V(\phi)$, it is possible to choose this function 
in order to produce a given $w_X(a)$ [or equivalently, $H(a)$]. 
For a flat universe, 
one can determine 
the function $V(\phi)$ 
for the quintessence model 
from the implicit relations \cite{em91}
\bear
  V(a) &=& \frac{H(a)}{16\pi G} [1-\Omega_{\rm kn}(a)] \nline
&\times&
\left[6H(a) + 2a H'(a) 
- \frac{a H(a) \Omega_{\rm kn}'(a)}{1-\Omega_{\rm kn}(a)}\right]
  \label{voft}
   \ear
    \bear
    \phi(a) &=&  \left[ \frac{1}{8\pi G}\right]^{1/2} 
\int \frac{\de a}{a} \nline
&\times&
     \left[ a\Omega_{\rm kn}'(a) - [1-\Omega_{\rm kn}(a)]
\frac{\de \ln H^2(a)}{\de \ln a}\right]^{1/2}
  \label{vphiquin}
    \ear
    Similarly, in the case of tachyonic scalar field, the potential function is given by \cite{padmanabhan02}
\bear
   V(a) &=& {3 H^2(a) \over 8\pi G}[1-\Omega_{\rm kn}(a)] \nline
&\times& 
\left\{ 1 + {2\over 3}{a H'(a)\over H(a)}-\frac{a\Omega_{\rm kn}'(a)}{3[1-\Omega_{\rm kn}(a)]}\right\}^{1/2}
   \label{vphitach}
   \ear
    \bear
  \phi(a) &=& \int \frac{\de a}{aH(a)} \nline
&\times&
\left\{\frac{a\Omega_{\rm kn}'(a)}{3[1-\Omega_{\rm kn}(a)]}
   -{2\over 3}{a H'(a)\over H(a)}\right\}^{1/2}
  \label{finalone}
  \ear
   
We can present an explicit example of a situation where 
$w_X$ is assumed to be known of the simple form
\be
w_X(z) = w_0 - w_1 (a - 1) = w_0 + w_1 \f{z}{1+z},
\label{wxz}
\e
where $w_0$ measures the current value of the parameter and $-w_1$ gives 
its rate of change (with respect to the scale factor) at the present 
epoch. In addition to simplicity, this 
parametrization has the advantage of giving 
finite $w_X$ in the entire range $0 < z < \infty$. 
We shall use this simple parametrization later to discuss 
the possibility of constraining $w_X$ from observations. 
The Hubble parameter in this case is given by
\bear
\f{H^2(z)}{H_0^2} &=& 
\Omega_m (1 + z)^3 \nline
&+& (1 - \Omega_m) (1 + z)^{3 (1 + w_0 + w_1)}
{\rm e}^{-3 w_1 z/(1 + z)}.
\ear
This equation of state parameter can be obtained 
(for small values of $|w_1|$)
from at least two scalar field Lagrangians -- those for the 
quintessence and tachyonic fields. The corresponding potentials 
$V(\phi)$ can be determined from 
equations (\ref{vphiquin}) and (\ref{vphitach}). Unfortunately, 
the relations cannot be written in a closed form 
(unless $w_1 = 0$) in this case. The form of the 
potential for both the cases can be calculated numerically for 
given values of $w_0$ and $w_1$, which are plotted in Figure~\ref{vphi}. 
The left frame shows $V(\phi)$ for the quintessence field while 
the right frame shows that for the and tachyonic field. 
The potentials are plotted for the parameter values 
$\Omega_m = 0.3, \Omega_X = 1 - \Omega_m = 0.7, 
w_0 = 0.5, w_1 = -0.1$ using equations 
(\ref{voft}) -- (\ref{finalone}). 
Since the equations give the potential $V(\phi)$ in the parametric form 
[$V(z)$, $\phi(z)$], each point on the $V - \phi$ curve can be labelled
by the corresponding value of redshift, $z$. 
We have pointed some particular redshifts in both the curves.
It is clear that the scalar field 
starts from a large value of the potential, rolls down as time 
progresses and the density contributed by the potential 
reaches a value $\sim \Omega_X$ at $z = 0$.

This discussion shows that 
even when $w_X(a)$ or $H(a)$ is known, it is 
{\em not} possible to proceed further and determine the nature of 
the unknown dark energy source using only kinematical
and geometrical measurements. 
This constitutes a serious theoretical problem in cosmology
if observations suggest that $w_X$ is not identically equal to $-1$
\cite{ht99}.

\section{Warm up: Current Supernova data and their analysis}
\label{sndata}

We shall next reanalyze the currently available supernova data in order to
stress certain features which are inherent in such an analysis. 
This may be considered
a warm up exercise for the next section.

The observations directly 
measure the apparent magnitude $m$ of a supernova and its 
redshift $z$. The apparent magnitude $m$ is related to the 
luminosity distance $d_L$ of the supernova through
\be
m(z) = M + 5 \log_{10} \left[\f{d_L(z)}{1~\mbox{Mpc}}\right] + 25,
\e
where $M$ is the absolute magnitude (which is believed to be constant 
for all supernovae of Type-Ia -- this is what is called the 
``standard candle hypothesis''). It is convenient to work with a 
dimensionless quantity (called the ``Hubble-constant-free'' luminosity 
distance)
\be
Q(z) \equiv \f{H_0}{c} d_L(z)
\e
which gives
\be
m(z) = {\cal M} + 5 \log_{10} Q(z),
\label{mq}
\e
where
\be
{\cal M} = M + 5 \log_{10} \left(\f{c/H_0}{1~\mbox{Mpc}}\right) + 25 
= M - 5 \log_{10} h + 42.38.
\e

Any model of cosmology will predict $Q(z)$ with some undetermined parameters 
(say, for example, $\Omega_m, \Omega_{\Lambda}$).
These parameters, along with ${\cal M}$ (which is 
itself a combination of the absolute magnitude of the Type-Ia supernova
and the Hubble constant),  
are obtained by comparing with observations. According to the standard 
hot big bang model of cosmology [see e.g., \citeN{padmanabhan02b}], 
\be
Q(z) = (1+z) \f{\sinh [x(z) \sqrt{\Omega_k}]}{\sqrt{\Omega_k}}, 
\e
where
\be
x(z) = \int_0^z \de z'\f{H_0}{H(z')}
\label{cosmoeq}
\e

The data in this paper is based on the 
redshift-magnitude relation of 54 Type-Ia supernovae 
[excluding 6 outliers
from the full sample of 60, which consists of 18 low-redshift 
points from the Cal\'{a}n/Tololo supernova survey \cite{hpssma96} and 
42 points from the Supernova Cosmology Project \cite{pag++99}] 
and that of supernova 1997ff at $z=1.755$ \cite{rng++01}.
The values of $m$ used 
have been already corrected for the supernova light curve width-luminosity 
relation, galactic extinction and possible K-correction. 
In addition, the magnitude 
for supernova 1997ff has been corrected 
for lensing effects \cite{brn++02}.
Given a cosmological model with some free parameters, we have obtained the 
best-fit parameter values and the corresponding covariance matrix 
using the Levenberg-Marquardt method \cite{ptvf92}.
The details of this fitting procedure, although quite well-known, are 
given in Appendix \ref{data_an} for completeness.

\begin{figure}
\begin{center}
\rotatebox{270}{\resizebox{0.45\textwidth}{!}{\includegraphics{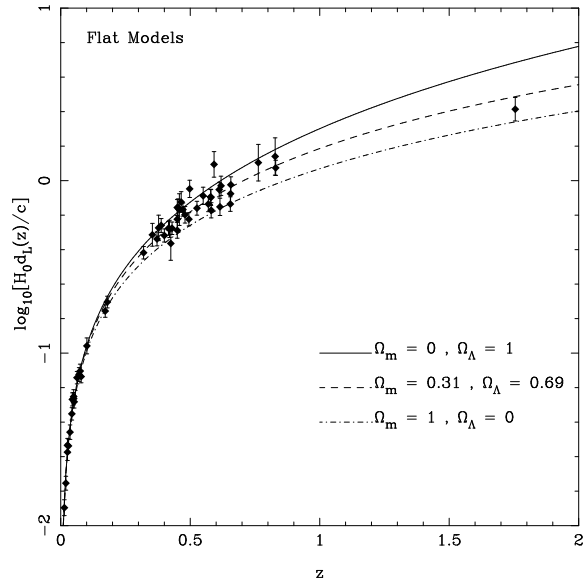}}}
\caption{Comparison between various flat models and 
the observational data. 
The observational data points, shown with error-bars, are obtained 
from Perlmutter et al. (1999) and Riess et al. (2001). In order to 
determine $Q(z)$ for these data points, we have used 
equation (\ref{mq}) with the best-fit value 
${\cal M} = 23.95$. 
}
\label{dataflat}
\end{center}
\end{figure}

Let us start with flat models with 
$\Omega_m + \Omega_{\Lambda} = 1; \Omega_k=0$. 
which are currently favoured strongly 
by CMBR data (for recent results, see \citeNP{sbc++02}). 
The simple analysis mentioned above gives 
a best-fit value of $\Omega_m$ (after marginalizing 
over ${\cal M}$) to be $0.31 \pm 0.08$. 
The best-fit ${\cal M}$ (after marginalizing 
over $\Omega_m$) is found to be $23.95 \pm 0.05$ 
(all the errors quoted in this paper are 1$\sigma$).
The comparison between three flat models and 
the observational data is shown in in Figure \ref{dataflat}. In order to 
determine $Q(z)$ for the observational data points, we have used 
equation (\ref{mq}) with the best-fit value 
${\cal M} = 23.95$. 

\begin{figure}
\begin{center}
\rotatebox{270}{\resizebox{0.45\textwidth}{!}{\includegraphics{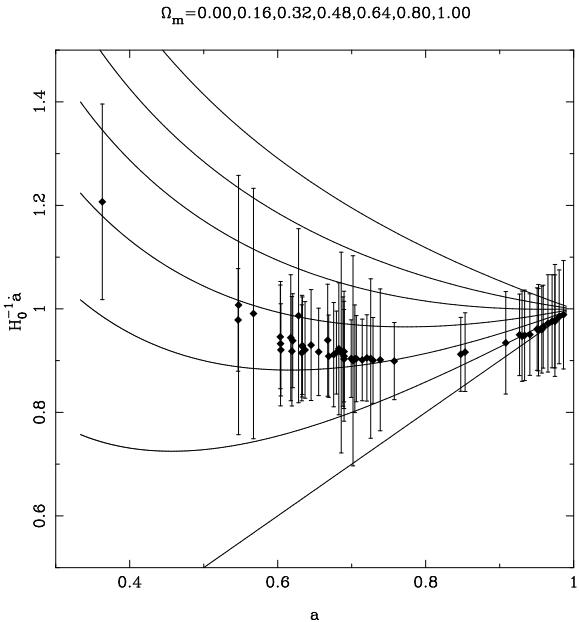}}}
\caption{The observed supernova data points in the $\dot{a} - a$ plane for 
flat models. The procedure for obtaining the data points and the 
corresponding error-bars are described in the text. The solid
curves, from bottom to top,  
are for flat cosmological models with 
$\Omega_m = 0.00, 0.16, 0.32, 0.48, 0.64, 0.80, 1.00$ respectively.
}
\label{adotomegam}
\end{center}
\end{figure}

Although the best-fit analysis shows that the data favour strongly 
for a positive non-zero $\Omega_{\Lambda}$ which, in turn, 
implies the presence of an accelerating universe, the same 
conclusion is not visually obvious 
from Figure \ref{dataflat}. 
In order to see that the data favours models with non-zero 
$\Omega_{\Lambda}$, one usually 
plots the supernova magnitude with respect to a fiducial 
best-fit model [see, for example, \citeN{pag++99}] -- however, 
to see the presence 
of an accelerating phase, it is more convenient to display 
the data as the phase portrait of the universe in the $\dot{a} - a$ plane. (The procedure for 
doing this is described in Appendix \ref{adot}; as far as we are aware, this has not been
done in literature before.)
The data points, with error-bars, in the $\dot{a} - a$ plane are shown in 
Figure \ref{adotomegam}. 
The solid curves plotted in Figure \ref{adotomegam} correspond 
to theoretical flat models with different $\Omega_m$.
The data points show a clear sign of an accelerating universe 
at low redshifts. Hence, in principle, one should be able to 
rule out non-accelerating models using only the low redshift data. 
However, as is clear from 
Figure \ref{adotomegam}, it is {\it not} possible 
to rule out any of the cosmological models using 
low redshift ($z \le 0.25$)
data because of large error-bars. (The fact that low redshift 
data cannot rule out any of  the cosmological models 
can be seen in Figure \ref{dataflat} also.)
On the other hand, high redshift data {\it alone} cannot 
be used to establish the existence of a cosmological constant. For example, 
one can use the freedom in the value of ${\cal M}$ to shift the 
data points vertically, and make them consistent with a 
decelerating model ($\Omega_m = 1$, topmost curve). It is the interplay 
between both the high and low redshift supernova data which leads 
to a clear indication of an accelerating phase.

All the above conclusions can be made more quantitative by studying 
at the confidence 
ellipses in the $\Omega_m - {\cal M}$ plane, shown in Figure \ref{fitflat}.
The method of drawing these ellipses is outlined in 
Appendix \ref{data_an}. 
\begin{figure*}
\begin{center}
\rotatebox{270}{\resizebox{0.33\textwidth}{!}{\includegraphics{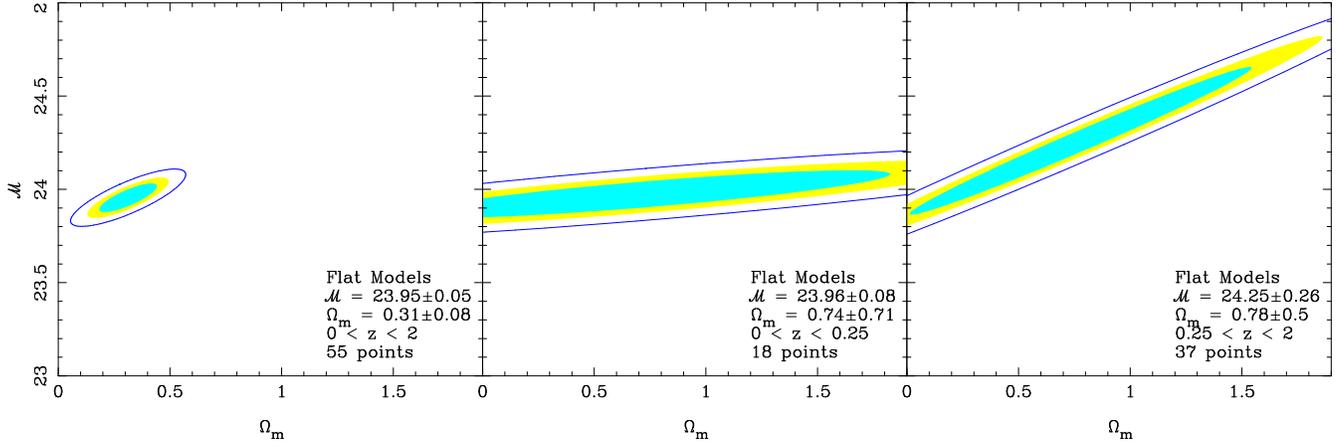}}}
\caption{Confidence region 
ellipses in the $\Omega_m - {\cal M}$ plane for flat models with 
non-relativistic matter and a cosmological constant. The 
ellipses corresponding to the 
68, 90 and 99 per cent confidence regions are shown.
In the left panel, all the 55 data points from Perlmutter et al. (1999) 
and Reiss et al. (2001) are used. 
In the middle panel, data 
points with $z < 0.25$ are used, while in the right panel, we have used  
data points with $z > 0.25$. We have indicated the best-fit values of 
$\Omega_m$ and ${\cal M}$ (with 1$\sigma$ errors).
}
\label{fitflat}
\end{center}
\end{figure*}
In the left panel, we have plotted the confidence regions using 
the full data set of 55 points. It is obvious that most of the 
probability is concentrated around the best-fit value.
The confidence contours in the middle and right panel are obtained by 
repeating the best-fit analysis for the low redshift data set 
($z < 0.25$) and high redshift data set ($z > 0.25$), respectively.
It is clear from the middle panel that, although the low redshift data 
can constrain ${\cal M}$ very well, it is unable to constrain 
$\Omega_m$. On the other hand, the high redshift data (right panel) 
is able to constrain neither ${\cal M}$ nor $\Omega_m$. 
In particular, the decelerating model ($\Omega_m=1$)
is quite consistent with both the low and high 
redshift data sets when 
they are treated separately. One needs to combine the 
low and high redshift data to constrain $\Omega_m$ -- because of the 
angular slant of the ellipses, a best-fit region around $\Omega_m = 0.31$
is isolated, as can be seen from the left panel.
[This analysis indirectly stresses the importance of any evolutionary effects.
If, for example, supernova at $z\gtrsim 0.25$ and supernova at
$z\lesssim 0.25$ have different absolute luminosities because of
some unknown effect, then the entire data set can be made consistent
with $\Omega_{\rm m} =1,\Omega_\Lambda =0$ model. It certainly
appears ad-hoc; but one should compare the ad-hocness in any of these
assumptions with the ad-hocness in 
introducing a dimensionless constant 
$\Lambda (G\hbar/c^3)
   \approx 10^{-123}$ in the physical system 
to explain the cosmological observations \cite{padmanabhan02c}!]

\begin{figure*}
\begin{center}
\rotatebox{270}{\resizebox{0.6\textwidth}{!}{\includegraphics{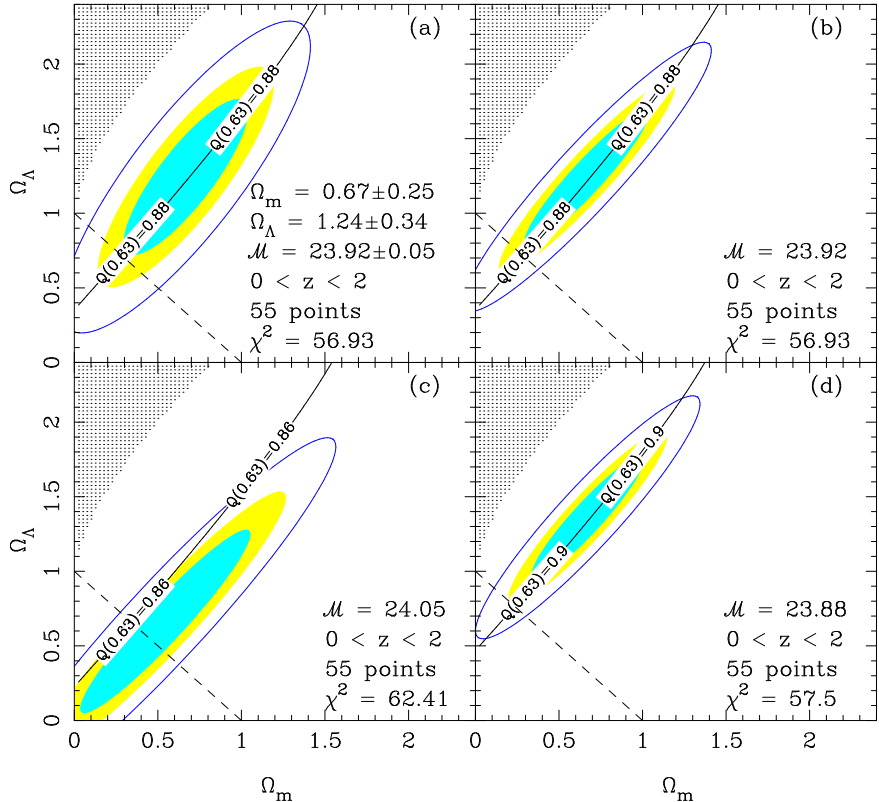}}}
\caption{Confidence region 
ellipses in the $\Omega_m - \Omega_{\Lambda}$ plane for models with 
non-relativistic matter and a cosmological constant. 
The ellipses corresponding to the 
68, 90 and 99 per cent confidence regions are shown.
(a) The confidence regions are obtained after marginalizing 
over ${\cal M}$. The best-fit values  of 
$\Omega_m$ and ${\cal M}$
are indicated (with 1$\sigma$ errors).
(b -- d) The confidence regions are obtained after fixing
${\cal M}$ to its best-fit mean value and the to the 
values in the wings of 1$\sigma$ limit respectively. 
The fixed value of ${\cal M}$ is indicated in the 
panel. The dashed line corresponds to the flat 
model ($\Omega_m + \Omega_{\Lambda} = 1$). 
We have indicated the value of $\chi^2$ for the best-fit parameters 
in all the panels.
The unbroken 
slanted line corresponds to the contour of 
constant luminosity distance, $Q(z = 0.63) =$ constant.
}
\label{fitallz}
\end{center}
\end{figure*}

We now generalize some of the results  for 
non-flat cosmologies. In this case, we have 
three free parameters, namely, $\Omega_m$, $\Omega_{\Lambda}$ and
${\cal M}$. 
The confidence region ellipses 
in the $\Omega_m$--$\Omega_{\Lambda}$ plane can be drawn 
in two ways -- (i) by keeping 
${\cal M}$ as a free parameter and then 
marginalize the joint probability distribution 
over ${\cal M}$ or, (ii) by fixing ${\cal M}$ to some constant value 
(and deal with only two free parameters, $\Omega_m$ and $\Omega_{\Lambda}$). 

The confidence region ellipses in the $\Omega_m$--$\Omega_{\Lambda}$
plane are shown in 
Figure \ref{fitallz} for the full data set. 
In panel (a) the confidence regions are obtained by marginalizing over 
${\cal M}$. The best-fit values  
are found to be $\Omega_m = 0.67 \pm 0.25$, 
$\Omega_{\Lambda} = 1.24 \pm 0.34$ and 
${\cal M} = 23.92 \pm 0.05$, as indicated 
in the panel. 
For panels (b -- d), the confidence contours are obtained by 
fixing ${\cal M}$ to a constant value rather than marginalizing over this 
parameter. The three frames correspond to 
the best-fit mean value and two values in the wings of 
1$\sigma$ from the mean, respectively. 
The fixed value of ${\cal M}$ is indicated in the 
panel. 

The main conclusions we can draw from this figure are:
(i) The results do not change significantly whether we marginalize 
over ${\cal M}$ or whether we use the best-fit value. 
One can see that the best-fit values of 
$\Omega_m$ and $\Omega_{\Lambda}$ do not change  
at all, it is only the the spread which decreases slightly  when we fix
${\cal M}$ to a particular value rather than marginalize it.
The two methods give such similar results because 
the probability of ${\cal M}$ is sharply peaked (the spread 
is $\sim 1$ per cent of the mean value).
(ii) The results exclude the SCDM model ($\Omega_m = 1, 
\Omega_{\Lambda} = 0$) at a high level 
of significance, taking into account the maximum uncertainty 
in ${\cal M}$.
The importance 
of this exercise will be evident later when we consider high and low redshift 
data points separately.
(iii) The slanted shape of the probability 
ellipses show that a particular linear combination of 
$\Omega_m$ and $\Omega_{\Lambda}$ is selected out by these observations
(which, in this case, 
turns out to be $0.82 \Omega_m - 0.57 \Omega_{\Lambda}$). 
This feature, of course, has nothing to do with supernova 
data and arises purely 
because the luminosity distance $Q$ depends strongly on  a 
particular linear combination of $\Omega_m$ and $\Omega_{\Lambda}$. 
This point is illustrated by plotting the contour of 
constant luminosity distance, $Q(z = 0.63) =$ constant. The coincidence 
of this line (which roughly corresponds to $Q$ at a redshift 
in the middle of the data) with the probability ellipses 
indicates that it is the dependence of the luminosity 
distance on cosmological parameters 
which essentially determines the nature of this result. 
This fact is known in the literature [see e.g., \citeN{gp95}], but 
we have not seen the actual data presented in this form. 
We shall use this result in the next section to discuss 
the possibility of determining the equation of state parameter 
for the evolving dark energy component. One should 
note that a complicated likelihood analysis may give 
confidence contours of a different shape (they are ellipses 
only because of the assumption of normal distribution of 
errors here) -- however, 
the shapes of those contours are bound show the degeneracy in 
$\Omega_m$ and $\Omega_{\Lambda}$ (arising purely 
from theory) mentioned above. We should also mention that 
the shapes of the contours of constant $Q$ at high redshifts 
are quite different from those at low redshifts -- hence more 
high redshift data can be useful in breaking the 
degeneracy between $\Omega_m$ and $\Omega_{\Lambda}$.

\begin{figure}
\begin{center}
\rotatebox{270}{\resizebox{0.45\textwidth}{!}{\includegraphics{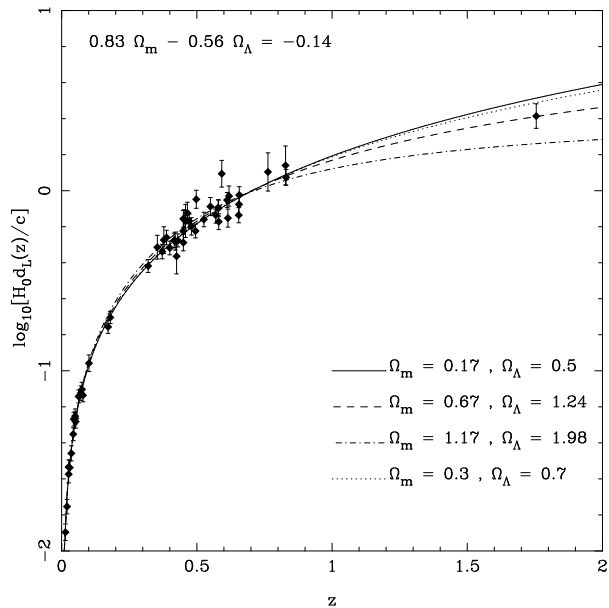}}}
\caption{The luminosity distance for a class of models with 
$\Omega_m$ varying between $0.67 \pm 0.50$ and $\Omega_{\Lambda}$
between $1.24 \pm 0.74$, but with a constant value 
of the combination $0.82 \Omega_m - 0.57 \Omega_{\Lambda}$. It is clear 
that when the combination $0.82 \Omega_m - 0.57 \Omega_{\Lambda}$ 
is fixed, low redshift observations cannot distinguish 
between the different models even if 
$\Omega_m$ and $\Omega_{\Lambda}$ vary significantly.
}
\label{qsens}
\end{center}
\end{figure}

As discussed in the third point above, the luminosity 
distance is sensitive only to 
a particular linear combination of $\Omega_m$ and $\Omega_{\Lambda}$, which 
is illustrated in Figure \ref{qsens}.
In this figure, $\Omega_m$ and $\Omega_{\Lambda}$ are treated 
as free parameters
(within $2\sigma$ bounds 
from the best-fit values mentioned above), but the combination 
$0.82 \Omega_m - 0.57 \Omega_{\Lambda}$ is held fixed. 
It turns out that $Q(z)$ is not very sensitive to 
individual values of $\Omega_m$ and $\Omega_{\Lambda}$ at low redshifts
when $0.82 \Omega_m - 0.57 \Omega_{\Lambda}$ is in the range 
$-0.14 \pm 0.11$. This is clear from 
Figure \ref{qsens} in which a wide variety of cosmological models 
are plotted along with the data for a constant value of the 
above combination. Though some of the models have unacceptable 
values of $\Omega_m$ and $\Omega_{\Lambda}$ 
(ruled out by observations of CMBR and structure formation), the supernova 
measurements alone cannot rule them out. Essentially, the data 
at $z < 1$ is degenerate on the linear combination 
$0.82 \Omega_m - 0.57 \Omega_{\Lambda}$. Our analysis of 
the supernova data shows that the best-fit value 
for this combination is 
$0.82 \Omega_m - 0.57 \Omega_{\Lambda} = -0.14 \pm 0.11$.

\begin{figure*}
\begin{center}
\rotatebox{270}{\resizebox{0.6\textwidth}{!}{\includegraphics{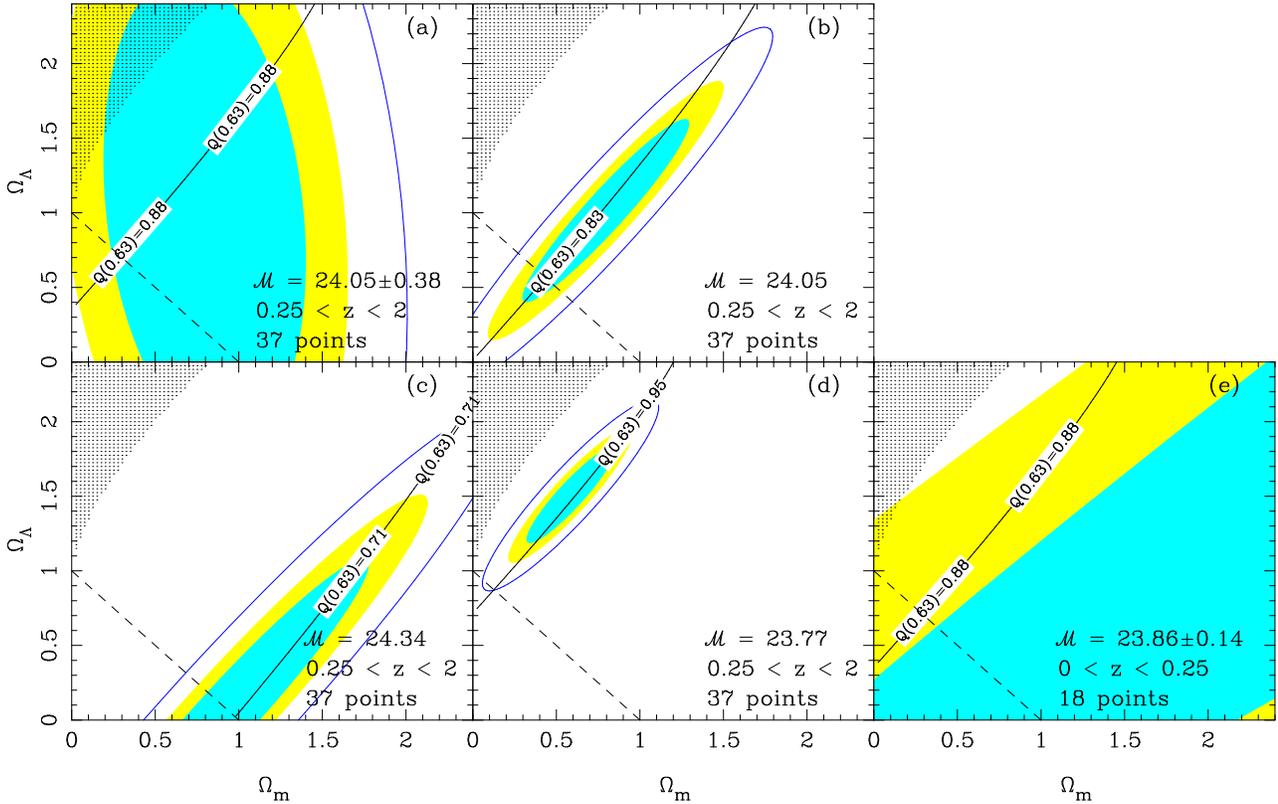}}}
\caption{Confidence region 
ellipses in the $\Omega_m - \Omega_{\Lambda}$ plane for models with 
non-relativistic matter and a cosmological constant. 
The ellipses corresponding to the 
68, 90 and 99 per cent confidence regions are shown.
(a) The confidence regions are obtained for 
high redshift points ($z > 0.25$), after marginalizing 
over ${\cal M}$. The best-fit value (with 1$\sigma$ error) of ${\cal M}$
is indicated. 
(b -- d) The confidence regions are obtained 
for high redshift points ($z > 0.25$), 
after fixing
${\cal M}$ to its best-fit value, the upper limit and the lower 
limit respectively. The fixed value of ${\cal M}$ is indicated in the 
panel.
(e) The confidence regions are obtained for 
low redshift points ($z < 0.25$), after marginalizing 
over ${\cal M}$. The best-fit value (with 1$\sigma$ error) of ${\cal M}$
is indicated. 
}
\label{fithighlowz}
\end{center}
\end{figure*}

Finally, we comment on the interplay between high and low 
redshift data for non-flat models. Just as in the case of the flat models, 
we divide the full data set into low ($z < 0.25$) and high 
($z > 0.25$) redshift subsets, and repeated the best-fit analysis. 
The resulting confidence contours are shown 
in Figure \ref{fithighlowz}. 
In panel (a), the confidence contours are plotted for the 
high redshift data set after marginalizing 
over ${\cal M}$.  As we stressed before, the SCDM model 
cannot be ruled out using high redshift data alone. 
In panels (b -- d), we show the corresponding results in which 
the values of ${\cal M}$ are fixed (rather than marginalizing 
over ${\cal M}$). Comparing the frames (a -- d) of Figure 
\ref{fithighlowz} with the frames (a -- d) of Figure 
\ref{fitallz} where the full data set was used, we can draw 
the following conclusions: (i) The best-fit value 
for ${\cal M}$ is now $24.05 \pm 0.38$; the $1\sigma$ error 
has now gone up by nearly an order of magnitude compared 
to the case where the full data was used. Because 
of this spread, the confidence contours are sensitive 
to the value of ${\cal M}$ one uses, unlike the situation 
where all the data points were used.
(ii) Our conclusions will now depend on the exact value of ${\cal M}$
used. For the best-fit mean value and the lower end of 
${\cal M}$, the high redshift data can rule out the SCDM model
[see frames (b) and (d)]. But for the higher end of the allowed 
$1\sigma$ range of ${\cal M}$, we cannot exclude 
the SCDM model [see frame (c)]. 
This essentially shows that it is difficult to 
rule out models when there are large 
uncertainties in ${\cal M}$.
Finally, we plot the contours for the 
low redshift data set after marginalizing 
over ${\cal M}$ in panel (e). It is clear that 
the low redshift data cannot be used to discriminate 
between cosmological models effectively. This is 
because $Q(z)$ at low redshifts is only very weakly dependent 
on the cosmological parameters. So, even though the acceleration 
of the universe is a low redshift phenomenon, we cannot reliably determine 
it using low redshift data alone.

\section{Constraints on evolving dark energy}
\label{evoldarken}

As we have seen in the previous two sections, supernova observations suggest 
the existence of a non-zero $\Omega_{\Lambda}$. However, this very 
existence of $\Omega_{\Lambda}$ raises serious theoretical 
problems [see \citeN{padmanabhan02c} for a detailed discussion]. These 
difficulties have led people to consider the possibility 
that the dark energy is not just a constant -- but is evolving 
with time. The evolution of the dark energy component can be 
parametrized by its equation of state parameter $w_X(z)$, the 
evolution of the density 
$\rho_X$ being given by equation (\ref{evolenergy}).
In this section, we shall examine the possibility of 
constraining $w_X(z)$ by comparing theoretical 
models with supernova observations. 

One simple, phenomenological, procedure for comparing observations 
with theory is to parametrize the function $w_X(z)$ in 
some suitable form and determine a finite set of parameters in this function 
using the observations. Theoretical models 
can then be reduced to a finite set 
of parameters which can be 
determined from observations. To illustrate 
this approach, let us 
assume that, in the flat universe, $w_X(z)$ is completely determined 
by its 
current value ($w_0$) and  
its rate of change with respect to the scale factor at the present 
epoch ($-w_1$)
and is given by 
the simple form as in equation (\ref{wxz}).

Since the cosmological constant corresponds 
to the case $w_0 = -1, w_1 = 0$, it 
is interesting to see the constraints on $w_0$ from supernova data even if we 
assume $w_1 = 0$. In such a study, it should be noted that (i) 
the acceleration of the universe requires $w_0 < -1/3$, and (ii)
all values of of $w_0$ other that $w_0 = -1$ leads to 
a dark energy density which evolves as $(1 + z)^{3 (1 + w_0)}$.
A simple best-fit analysis shows that for a flat model with 
$\Omega_m = 0.31$ and ${\cal M} = 23.95$ (the best-fit 
parameters for flat models, obtained in the previous section), 
the best-fit value of 
$w_0$ is $-1.01 \pm 0.12$ (which is nothing but the 
conventional cosmological constant). The data clearly rules out 
models with $w_0 > -1/3$ at a high 
significance level, thereby supporting the existence of 
a dark energy component with negative pressure.

\begin{figure*}
\begin{center}
\rotatebox{270}{\resizebox{0.33\textwidth}{!}{\includegraphics{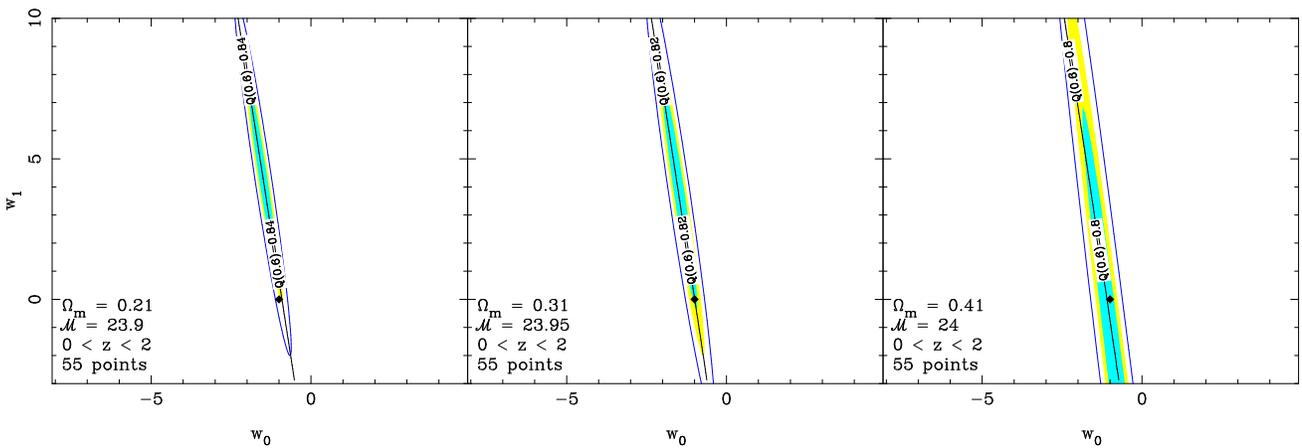}}}
\caption{Confidence region 
ellipses in the $w_0 - w_1$ plane for flat models with 
$\Omega_m = 0.21, 0.31$ and 0.41 respectively, as indicated 
in the frames. 
The value of ${\cal M}$ is chosen to be the best-fit value, 
which is also indicated. The 
ellipses corresponding to the 
68, 90 and 99 per cent confidence regions are shown.
The square point denotes the equation of state 
for a universe with a non-evolving dark energy component (the 
cosmological constant). The unbroken 
slanted line corresponds to the contour of 
constant luminosity distance, $Q(z = 0.6) =$ constant.
}
\label{fitw0w1}
\end{center}
\end{figure*}

One can extend the analysis to find the constraints in the 
$w_0 - w_1$ plane. 
[There is extensive literature on determining 
dark energy parameters from the current and proposed 
future supernova observations, e.g., see 
\citeN{mbms02}; \citeN{wa02}; \citeN{ge02} and references therein.]
As before, we assume a flat universe 
with a fixed value of $\Omega_m$ in the range (0.21 -- 0.41) and ${\cal M}$ is 
fixed to the corresponding best-fit value. The 
confidence contours for the three models are shown in Figure \ref{fitw0w1}.
\begin{figure*}
\begin{center}
\rotatebox{270}{\resizebox{0.6\textwidth}{!}{\includegraphics{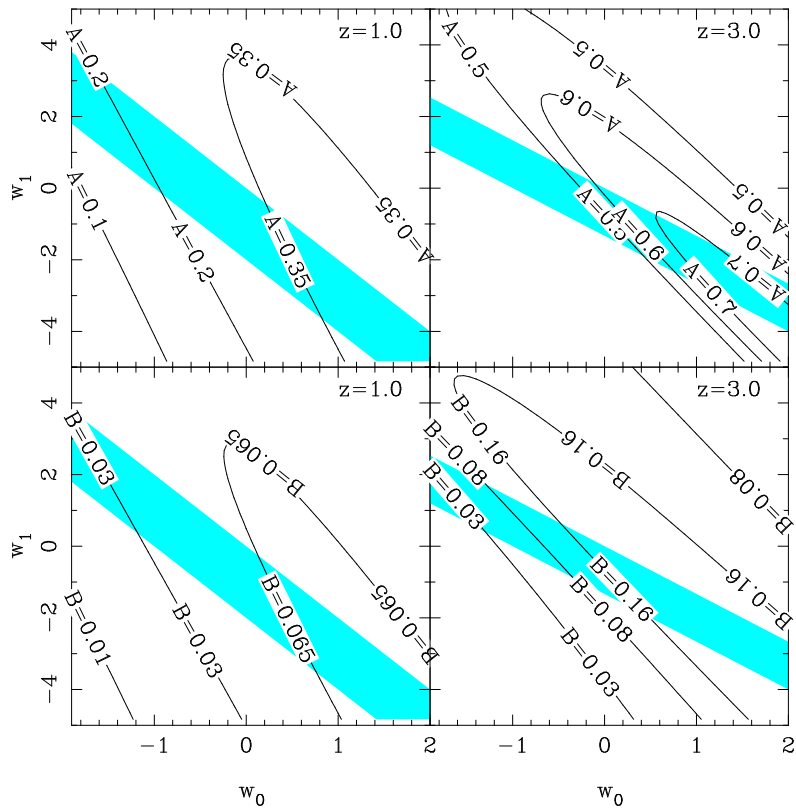}}}
\caption{Sensitivity of $Q(z)$ to the 
parameters $w_0$ and $w_1$. The curves 
correspond to the constant values for the fractional change in $Q$ for unit 
change in $w_0$ (top frames) and $w_1$ (bottom frames) 
for two redshifts $z = 1$ (left frames) and 
$z = 3$ (right frames). The value of 
$Q$ for a particular curve is indicated. A flat cosmological model 
with $\Omega_m = 0.31$ has been used.
The shaded bands across the frames correspond to the regions in 
which $-1 \le w_X(z) \le 0$.
}
\label{ab}
\end{center}
\end{figure*}
\begin{figure*}
\begin{center}
\rotatebox{270}{\resizebox{0.6\textwidth}{!}{\includegraphics{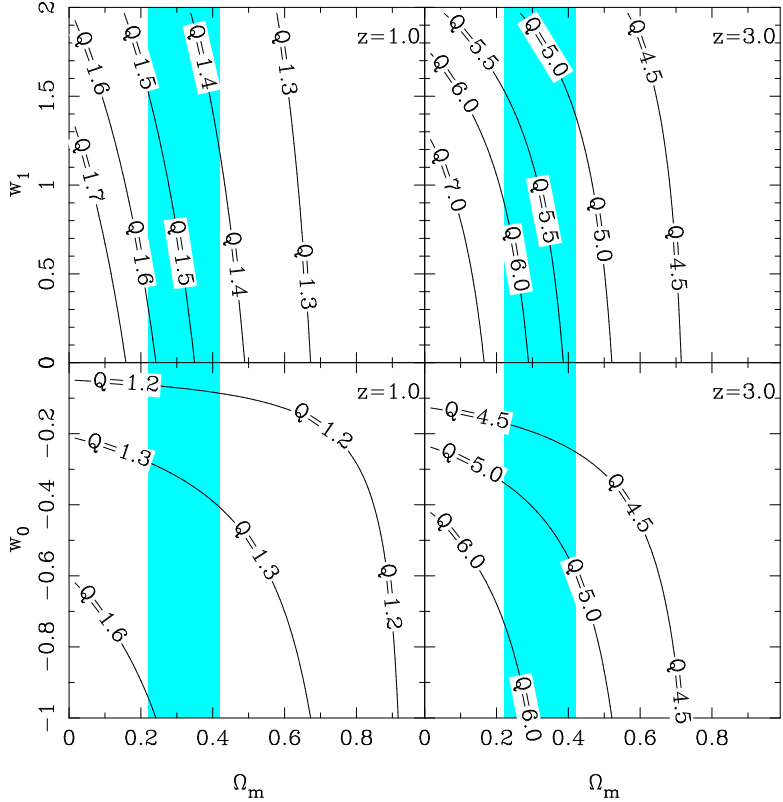}}}
\caption{Contours of constant $Q$ in the 
$\Omega_m - w_0$ (bottom frames) and $\Omega_m - w_1$ (top 
frames) planes at two redshifts $z = 1$ (left frames) and 
$z = 3$ (right frames). The value of $w_0$ is fixed 
to $-1$ in the top frames, while $w_1$ is fixed to 0 in 
the bottom frames. 
}
\label{q}
\end{center}
\end{figure*}
The square point denotes the equation of state 
for a universe with a non-evolving dark energy component (the 
cosmological constant). 
The main points revealed by this figure are:
(i) The equation of state corresponding to the 
cosmological constant is within the 
$1\sigma$ contour for $0.21 < \Omega_m < 0.41$. 
(ii) Despite 
the uncertainties in $w_1$, models with 
$w_0 > -1/3$ are ruled out at high significance level for 
$\Omega_m < 0.4$. (One can accommodate 
models with $w_0 > -1/3$ only for $\Omega_m > 0.4$ for 
very high negative values of $w_1$.)
(iii) The shape of the confidence 
contours clearly indicates that the data is not as sensitive to 
$w_1$ as compared to $w_0$. 
Since the supernova observations essentially measure 
$Q(z)$, this is a clear indication that $Q(z)$ is 
comparatively 
insensitive to $w_1$. To see this, note that in Figure \ref{fitw0w1}, 
the position of the 
slanted line (which corresponds to $Q(z = 0.6) =$ constant)
exactly coincides with the contour ellipses. 
We mention again that the shapes of the confidence contours 
may not remain 
as simple ellipses if one carries out a complicated likelihood
analysis -- however, as our conclusions are based 
on the sensitivity of $Q(z)$ on $w_0$ and $w_1$, they are likely to 
remain  the same for any analysis.
Since the determination of 
$w_0$ and $w_1$ is of considerable 
importance, we shall provide an 
illustration of the relative sensitivities 
of geometrical features to $w_0$ and $w_1$.

As we have stressed in the previous section and also 
above, the supernova observations essentially measure 
$Q(z)$, and hence
the accuracy in the determination of $w_0$ and $w_1$ from 
(both present and planned future) supernova observations will crucially
depend on how sensitive $Q$ is to the changes in $w_0$ and $w_1$ 
[which is clear from the coincidence 
of the $Q = $ constant line with the probability ellipses
in Figure \ref{fitw0w1}].
A good measure of the sensitivity is provided by the two parameters 
\bear
A(z; w_0, w_1) &\equiv& \f{\de}{\de w_0} \ln[Q(z; w_0, w_1)], \nline
B(z; w_0, w_1) &\equiv& \f{\de}{\de w_1} \ln[Q(z; w_0, w_1)].
\label{abdef}
\ear
Like $Q(z; w_0, w_1)$, the parameters $A$ and $B$ can be computed 
for a particular cosmological model 
in a straightforward manner. At any given redshift $z$, 
we can plot contours of constant $A$ and $B$ in the 
$w_0 - w_1$ plane. Figure \ref{ab} shows the result 
of such analysis for flat models with $\Omega_m = 0.31$.
The two frames on the left are for $z = 1$, while those on the right 
are for $z = 3$. The top frames show the contours of constant 
$A$ while the bottom frames show the contours of constant 
$B$.
From the definitions (\ref{abdef}) it is clear that $A$ and $B$ 
can be interpreted as the fractional change in $Q$ for unit change 
in $w_0$ and $w_1$, respectively. For example, along the 
line marked $A = 0.2$ (in the top left frame), $Q$ will 
change by 20 per cent for unit change in $w_0$. It is clear from 
the two top frames that for most of the interesting region 
in the $w_0 - w_1$ plane, changing $w_0$ by unity changes 
$Q$ by about 10 per cent or more. Comparison of 
the two cases $z = 1$ and $z = 3$ (the two top frames) shows 
that the sensitivity is higher at high redshift, as to be expected.
The shaded bands across the frames correspond to the regions in 
which $-1 \le w_X(z) \le 0$, which is of 
primary interest in 
constraining dark energy with negative pressure. One concludes 
from the above discussion that determining 
$w_0$ from $Q$ fairly accurately will not be too daunting 
a task.

The situation, however, is quite different for $w_1$ as
illustrated in the two bottom frames. For the same region 
of the $w_0 - w_1$ plane, $Q$ changes only by a few per cent 
when $w_1$ changes by unity. This means that $Q$ is much less 
sensitive to $w_1$ than to $w_0$. It is going to be 
significantly more difficult to determine a value for 
$w_1$ from observations of $Q$ (like supernova observations) 
in the near future. Comparison of the 
two cases $z = 1$ and $z = 3$ again shows that the sensitivity 
is somewhat better at high redshifts, but only marginally.

In the above analysis, we have treated $\Omega_m$ to be a 
constant. The situation is made worse by the fact that 
$Q$ also depends on $\Omega_m$. If the variation 
of $\Omega_m$ mimics that of $w_0$ or $w_1$, then one also 
needs to determine the 
sensitivity of $Q$ to $\Omega_m$. 
Figure \ref{q} shows contours of 
constant $Q$ in the $\Omega_m - w_1$ (top frames) and 
the $\Omega_m - w_0$ (bottom frames) planes at 
two redshifts $z = 1$ (left frames) and $z = 3$ (right frames). 
The two top frames 
show that if one varies the value of $\Omega_m$ in the allowed range, say, 
(0.2 -- 0.4), one can move along the curve of constant $Q$ and 
induce a fairly large variation ($\sim$ 1 -- 2) in $w_1$. In other 
words, large changes in $w_1$ can be easily 
compensated 
by small changes in $\Omega_m$ while maintaining 
the same value for $Q$ at a given redshift. This shows that the 
uncertainty in $\Omega_m$ introduces further difficulties in determining 
$w_1$ accurately from measurements of $Q$.
The two bottom frames show that the situation is better for $w_0$. 
The curves are much less steep and 
hence varying 
$\Omega_m$ does not 
induce large variations in $w_0$.
We once again are led to the conclusion that unambiguous 
determination of $w_1$ from data will be quite difficult.
It appears that supernova observations may not be of great 
help in ruling out (non-evolving) cosmological constant as the 
major dark energy component.

\section{Discussion}

The supernova data on the whole,  
rules out non-accelerating models with very high significance
level, 
as noted and stressed by various authors \cite{rfc++98,pag++99,riess00}. 
It is thus more or less conclusive that we have to 
look for some form of matter which has negative pressure.
However, it is interesting 
to note that if we divide the data set into high and low redshift subsets, 
neither 
of the subsets are able to rule out the decelerating models -- it is only 
the interplay between the high and low redshift data which implies the 
result quoted above. This analysis indirectly stresses the 
importance of any evolutionary effects. Any possible evolution 
in the absolute magnitude of the supernovae, if detected, 
might allow the decelerating models to be consistent with 
the data. However, our understanding of the 
complicated physical processes in the supernovae is 
definitely not well enough to discuss the possibility 
of detecting supernova evolution in near future. 
Although the intrinsic evolution of the SN is 
an interesting problem, 
addressing these issues is beyond the 
scope of this paper.

The supernova data available today is not enough to fix the parameters 
$\Omega_m$ and $\Omega_{\Lambda}$ independently. Instead, a
particular combination of the two parameters is chosen out by the data. 
This degeneracy 
has to do with the dependence of the luminosity distance on these 
parameters. Since 
supernova observations essentially measure the luminosity distance
$Q(z)$, the accuracy in the determination of various
parameters from 
supernova observations will crucially
depend on how sensitive $Q$ is to the changes in those parameters. 
This analysis can be done entirely from theory 
[using Fisher information matrix; see for example 
\citeN{efstathiou99}] and, 
in principle, can provide 
nice handle on which combinations of 
parameters are expected to be constrained  
from the supernova data. 

The key issue regarding dark energy is whether it 
is a constant or whether it is evolving with time. In particular, 
one is interested to determine the evolution of the equation of 
state, $w_X$ of the dark energy component. 
To address this question, we have 
taken a simple phenomenological model 
for $w_X$ having the form $w_X(a) = w_0 - w_1 (a - 1)$. 
The sensitivity of the luminosity distance (and hence, the supernova data) 
on $w_0$ and $w_1$ 
can be measured by 
the two parameters ($A$ and $B$) introduced in the paper, 
which are essentially  the fractional 
changes in $Q(z)$ for unit change in 
$w_0$ and $w_1$, respectively. The parameters 
$A$ and $B$ can be obtained entirely from theory and the conclusions 
drawn from  them 
are valid for current as well as future supernova observations. We find 
that $Q(z)$ is quite sensitive 
to $w_0$ -- hence, one can constrain the current 
value of $w_X$ quite well. However, 
$Q(z)$ is comparatively insensitive 
to $w_1$, thus determining the 
evolution of $w_X$ will be a difficult task. The situation 
is further worsened when we take the uncertainties in $\Omega_m$ 
into account. Our results are consistent with other statistical analyses 
done by  
\citeN{astier00}, \citeN{mbs01}, \citeN{wa01}, \citeN{mbms02}, \citeN{wa02}. 
They have shown that it is  
possible to constrain the present value of $w_X$ with future 
missions like ``SuperNovae Acceleration Probe'' (SNAP); however, constraining 
the evolution of $w_X$ will not be easy using SNAP, particularly when 
$\Omega_m$ is not known to a high accuracy.

\section*{Acknowledgements}
T.R.C. is supported by the University Grants Commission, India.

\bibliography{mnrasmnemonic,astropap}
 
\bibliographystyle{mnras}

 

\appendix

\section{Data Analysis}
\label{data_an}

In this section, we shall outline the method used for obtaining the 
best-fit (theoretical) 
parameters from observational data, and draw the corresponding 
confidence contours. For simplicity, we shall assume that the measurement 
errors are normally distributed. 
Suppose we have $M$ observational data points 
denoted by $m_{\rm obs}(z_i); i=1,...,M$ with corresponding errors 
$\sigma_m(z_i)$. For a given theoretical model 
$m_{\rm th}(z; c_{\alpha})$ with $\nu$ free parameters 
$c_{\alpha}; \alpha = 1,...,\nu$, one can construct the 
quantity
\be
\chi^2(c_{\alpha}) = 
\sum_{i=1}^M \left[
\f{m_{\rm obs}(z_i) - m_{\rm th}(z_i; c_{\alpha})}{\sigma_m(z_i)}
\right]^2
\e
The best-fit parameters $\bar{c}_{\alpha}$ are obtained by minimizing the 
above quantity 
\be
\left[
\f{\partial \chi^2}{\partial c_{\alpha}}
\right]_{c_{\alpha} = \bar{c}_{\alpha}} = 0
\e
using the Levenberg-Marquardt method \cite{ptvf92}.
One can show that the quantity 
$\Delta \chi^2 = \chi^2(c_{\alpha}) - 
\chi^2(\bar{c}_{\alpha})$ follows a chi-square 
distribution with $\nu$ degrees of freedom.

To obtain the confidence intervals, let us first define the curvature matrix 
\be
A_{\alpha \beta} = \f{1}{2} \left[
\f{\partial^2 \chi^2}
{\partial c_{\alpha} \partial c_{\beta}} 
\right]_{c_{\alpha} = \bar{c}_{\alpha}}.
\e
The covariance matrix $C_{\alpha \beta}$ is simply the
inverse of the curvature matrix. The probability 
distribution of the parameters is given by
\be
{\cal P}(c_{\alpha}) = {\rm const} \times \exp\left[-\f{1}{2}
\sum_{\alpha,\beta=1}^{\nu}(c_{\alpha} - \bar{c}_{\alpha}) 
A_{\alpha \beta} (c_{\beta} - \bar{c}_{\beta}) \right]
\label{probdist}
\e

Suppose we are interested in the confidence regions for 
a subset of, say, $\nu'$ parameters ($\nu' \le \nu$). In that case, the 
the regions 
are obtained by simply marginalizing the 
probability distribution over the rest $\nu - \nu'$ parameters 
[which is equivalent to integrating the probability 
distribution (\ref{probdist}) over the rest $\nu - \nu'$ parameters]. 
This can be done simply by taking the 
full $\nu \times \nu$ covariance matrix $C_{\alpha \beta}$ and 
copying the intersection of the 
$\nu'$ rows and columns 
corresponding to the parameters of interest into a 
$\nu' \times \nu'$ matrix $C'_{\alpha \beta}$. The 
inverse of this will give the corresponding curvature matrix 
$A'_{\alpha \beta}$ \cite{ptvf92}. The marginalized 
probability distribution will be simply
\be
{\cal P'}(c_{\alpha}) = {\rm const} \times \exp\left[-\f{1}{2}
\sum_{\alpha,\beta=1}^{\nu'}(c_{\alpha} - \bar{c}_{\alpha}) 
A'_{\alpha \beta} (c_{\beta} - \bar{c}_{\beta}) \right]
\e

Next, one needs to find the quantity 
$\Delta \chi^2_p$, such that the probability 
of a chi-square variable with $\nu'$ degrees of freedom 
being less than $\Delta \chi^2_p$ is $p$, where
$p$ is  the desired confidence limit (e.g., 0.68 or 0.95). This can be 
obtained by solving the equation
\be
p = \f{1}{2^{\nu'/2} \Gamma(\nu'/2)}
\int_0^{\Delta \chi^2_p} \de u ~ u^{(\nu'/2) - 1} {\rm e}^{-u/2}
\e
where $\Gamma(z)$ is the gamma function.
For a given $p$ and 
a set of $\nu'$ parameters, the equation for the boundary 
of the confidence region in the $\nu'$-dimensional 
space is given by the equation
\be
\Delta \chi^2_p = \sum_{\alpha,\beta=1}^{\nu'}(c_{\alpha} - \bar{c}_{\alpha}) 
A'_{\alpha \beta} (c_{\beta} - \bar{c}_{\beta}) 
\e

\section{Determination of the Hubble parameter $H(\lowercase{z})$ from supernova observations}
\label{adot}

\begin{figure}
\begin{center}
\rotatebox{270}{\resizebox{0.45\textwidth}{!}{\includegraphics{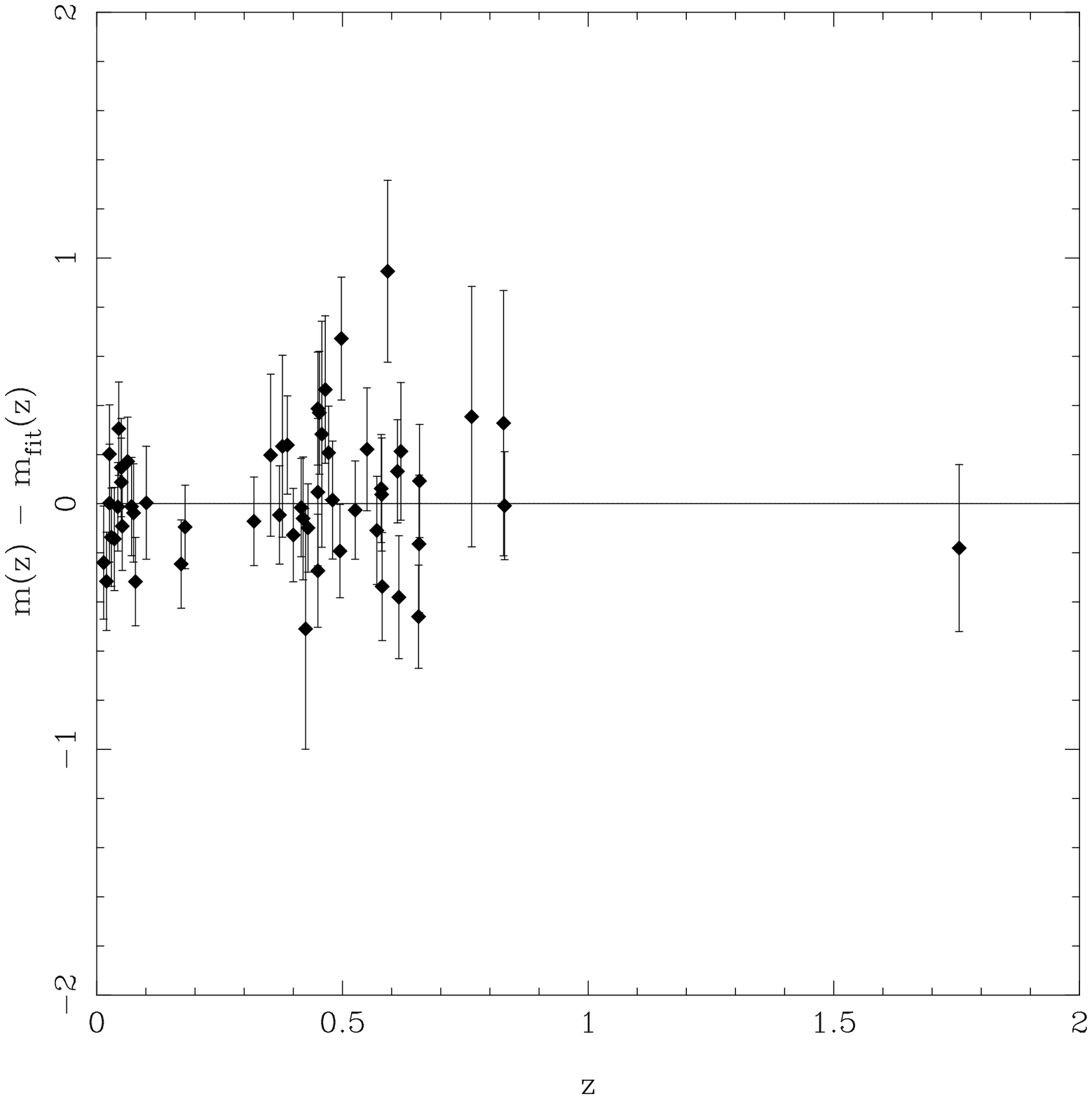}}}
\caption{The difference between the actual 
data points and the fit (\ref{fitfunc}) 
to the data.
}
\label{mfit}
\end{center}
\end{figure}

The observational data used in this paper can be fitted 
by the function of simple form
\be
m_{\rm fit}(z) 
= a_1 + 5 \log_{10} \left[\f{z (1 + a_2 z)}{1 + a_3 z}\right].
\label{fitfunc}
\e
where the parameters are
\be
a_1 = 23.95 \pm 0.05;
a_2 = 2.00 \pm 1.18;
a_3 = 1.03 \pm 0.88
\e
The $\chi^2$ per degree of freedom for the best-fit values of the 
parameters 
is found to be 1.05, which shows the fit is reasonably good. 
The difference between the data points and the fit is shown in Figure 
\ref{mfit}. 
We can then represent the luminosity distance obtained 
from the data by the function
\be
Q_{\rm fit}(z) = 10^{0.2 [m_{\rm fit}(z) - {\cal M}]}
\e
Note that one needs to fix the value of ${\cal M}$ to 
obtain the function $Q_{\rm fit}(z)$.
For flat models, the value of ${\cal M}$ is well-constrained 
(the error being $\sim 1$ per cent) -- hence we can use 
the best-fit value of ${\cal M} = 23.95$ to obtain
\be
Q_{\rm fit}(z) = \f{z (1 + 2.00 z)}{1 + 1.03 z}
\e
For flat models, it straightforward to obtain the Hubble parameter
from $Q(z)$. 
In particular, we are interested in the quantity
\be
H_0^{-1} \dot{a}(z) = \left[(1+z) \f{\de}{\de z} \left\{\f{Q(z)}{1+z}\right\}
\right]^{-1}
\e
which will enable us to plot the data points in the 
$\dot{a} - a$ plane. The determination of the corresponding 
error-bars is a non-trivial exercise. In this paper, the errors are 
obtained from the relation
\be
H_0^{-1} \dot{a}(z)(1 + \epsilon_{\dot{a}}(z)) = 
\left[(1+z) \f{\de}{\de z} \left\{\f{Q(z)(1 + \epsilon_{Q}(z))}{1+z}\right\}
\right]^{-1},
\e
where $\epsilon_{\dot{a}}$ and $\epsilon_{Q}$ denote the fractional error
in ${\dot{a}}$ and $Q$, respectively. 
From equation (\ref{mq}), we have 
$\epsilon_{Q}(z) = 0.2 \ln 10 ~ \sigma_m(z)$, where 
$\sigma_m(z)$ is the total uncertainty in the observed 
magnitude. Since there 
is no systematic evolution of the observed $\sigma_m(z)$ (see 
Figure \ref{mfit}), we have, to 
the lowest approximation, the fractional error in $\dot{a}$ 
\be
\epsilon_{\dot{a}}(z) = \epsilon_{Q}(z) = 0.2 \ln 10 ~ \sigma_m(z).
\e
Thus, one can plot the quantity $\dot{a}(z)$ for the 
observational data with error-bars given by the above equation.

Note that the analysis assumes that 
the observed errors in the measurement of $z$ (or, equivalently $a$) 
are negligible. Strictly speaking, the errors in $\dot{a}$ calculated 
above are strict lower limits -- they can be slightly 
higher because of the errors arising from (i) the fitting function and 
(ii) any systematic evolution of the observed $\sigma_m(z)$. However, 
these effects are unlikely to affect any of the conclusions we 
have drawn in this paper.

\end{document}